\documentclass[
    ,final            
  ]
  {aipproc}

\layoutstyle{6x9}

\usepackage{epstopdf}
\usepackage{wasysym}

\newcommand{\msun}{M_{\odot}}
\newcommand{\mwimp}{m_\chi}

\begin{document}

\title{Supermassive Dark Stars: Detectable in JWST and HST}


\author{Katherine Freese}{
  address={Michigan Center for Theoretical Physics, Department of Physics, University of Michigan, Ann Arbor, MI 48109}
}
\author{Eduardo Ruiz}{
  address={Michigan Center for Theoretical Physics, Department of Physics, University of Michigan, Ann Arbor, MI 48109}
}
\author{Monica Valluri}{
  address={Department of Astronomy, University of Michigan, Ann Arbor, MI 48109}
}
\author{Cosmin Ilie}{
  address={Michigan Center for Theoretical Physics, Department of Physics, University of Michigan, Ann Arbor, MI 48109}
}
\author{Douglas Spolyar}{
  address={Center for Particle Astrophysics, Fermi National Accelerator Laboratory, Batavia, IL, 60510}
}
\author{Peter Bodenheimer}{
  address={UCO/Lick Observatory, Dept. of Astronomy and Astrophysics, University of California, Santa Cruz, CA 95064}
}

\begin{abstract}
 The first stars to form in the history of the universe may have been powered by dark matter annihilation rather than by fusion.  This new phase of stellar evolution may have lasted millions to billions of years. These dark stars can grow to be very large, $>10^5 \msun$, and are relatively cool ($\sim 10^4$ K).  They are also very bright, being potentially detectable in the upcoming James Webb Space Telescope or even the Hubble Space Telescope.  Once the dark matter runs out, the dark stars have a short fusion phase, before collapsing into black holes (BH).  The resulting BH  could serve as seeds for the (unexplained)  supermassive black holes at high redshift and at the centers of galaxies.
\end{abstract}

\maketitle


\section{Introduction}

The first stars form at $z>10$ at the centers of dark matter (DM) haloes of  $10^6 - 10^8 \msun$ (Barkana \& Loeb
(2001); Yoshida et al. (2003); Bromm \& Larson (2004); and Ripamonti\& Abel (2005)).  The DM inside the stars may provide a power source for these stars, resulting in Dark Stars (DS) (Spolyar,
Freese, and Gondolo 2008) that are powered by  DM annihilation rather by fusion.  This first phase of stellar evolution can be quite long, lasting millions to billions of years.  

The identity of DM is still unknown.   Weakly interacting massive particles (WIMPs) are the best motivated DM candidates, since the WIMP cross section yields the correct relic density necessary to explain dark matter abundance. Typical supersymmetric WIMPs are their own antiparticles, 
and thus annihilate among themselves wherever the WIMP density is high enough.
The WIMP mass ranges from 1 GeV - 10 TeV (our canonical value is 100 GeV), and  the standard annihilation cross section is
\begin{equation}
\langle \sigma v \rangle_{ann} = 3 \times 10^{-26} {\rm cm}^3 / {\rm sec} .
\label{eq:sigmav}
\end{equation}
It is this same annihilation process that took place in the early universe, that is also the source
of WIMP heating for the first stars.

Dark matter annihilation is particularly powerful in the first stars,
 as they formed in a very high DM density environment: they formed at the right place (at the high density centers of haloes) and at the right time (at high redshift, with density scaling as $(1+z)^3$).

A protostellar cloud made of primordial hydrogen and helium forms at the center of the DM minihalo; as the gas cools and collapses, the cloud compresses the DM, leading to an increase in dark matter annihilation rates.  At higher densities, DM heating overwhelms any cooling mechanism, stopping the cloud from collapsing further, and a DS forms.   We have computed the stellar structure and evolution of the DS,
and find that they are cool ($\sim 10^4$K), puffy (radii $\sim$ A.U.), and massive (Freese et al. 2008a).  As long as
DM powers the star, the DS remains cool enough to avoid producing photons hot enough to prevent further accretion onto the star. The DS continues to grow, even becoming Supermassive
Dark Stars (SMDS) with $M_*> 10^5 \msun$. 

This talk reports on the work of Freese et al (2010)
and in addition discusses detectability in Hubble Space Telescope (HST).
Various other authors have explored the repercussions of DM heating
in the first stars including Taoso et al. 2008; Yoon et al. 2008; Ripamonti et al. 2009; Iocco et al. 2008; Schleicher et al.(2008, 2009); Ripamonti et al (2009); Ripamonti et al (2010); Umeda et al (2009); Sivertsson and Gondolo (2010); Zackrisson et al (2010a) as well as additional references discussed in the text.

We also mention that there have been several hints of WIMP DM detection recently.  The direct detection experiments DAMA (Bernabei et al 2009) and COGENT (Collar \& McKinsey (2010)); PAMELA positron excess (Adriani et al (2009)); gamma rays detected by FERMI (Abdo et al (2009)); 
and the WMAP haze (Hooper, Finkbeiner, and Dobler (2007)) all have anomalous events that may originate from DM in the Galaxy.  The feeling in the community is that discovery is near. 

\section{Dark Stars}
 The DM heating rate is given by
\begin{equation}
\Gamma_{DMHeating} = f_Q n_\chi^2 \langle \sigma v\rangle_{ann} \mwimp =  f_Q \rho_x^2\langle\sigma v\rangle_{ann} /\mwimp.
\label{eq:Q}
\end{equation}
Here $n_\chi$ is the DM density, $\rho_\chi$ is the DM mass density, and  $f_Q$ is the fraction of annihilation energy that is trapped by the star and goes into heating it.
The final products of the annihilation are typically of order
1/3 electron/positron, 1/3 gamma rays, and 1/3 neutrinos; these particles typically 
have energies roughly about 1/10 of the WIMP mass.  Since neutrinos simply escape from the
star, we take $f_Q = 2/3$ as our fiducial value.

 Three conditions are required for the formation of a DS, (1) a sufficiently high DM density, (2) the annihilation products get trapped in the DS, and (3) DM heating overpowers $H_2$ cooling.  
As mentioned previously, the first stars already have relatively high DM density because they
form at the centers of haloes at high redshift.  Then, the density can be further enhanced by
two mechanisms: adiabatic contraction (AC) (Blumenthal et al. 1985; Freese et al (2009)) and capture
(Iocco (2008); Freese, Spolyar, Aguirre (2008)).  AC takes place already in the early stages of the collapsing protostellar
cloud.  The collapsing gas provides a growing gravitational potential to pull in more DM, enhancing its density and the 
resulting annihilation. Second, the annihilation products must get trapped within the DS; this
criterion is satisfied once the collapsing cloud reaches sufficiently high density.  Third,
the DM heating must overpower molecular hydrogen cooling.  For a 100 GeV WIMP, this
happens at a gas density $n \sim 10^{13}$ cm$^{-3}$ (Spolyar et al. (2008)).  The star collapses a little further\footnote{We note that we are in agreement with Ripamonti et al (2010) in that the star does have to collapse
a little further before equilibrium is established; they interpret this a possible failure of the 
ability of the DM to halt the collapse, whereas we perceive this to be the approach to 
equilibrium.} in order to 
establish thermal and hydrostatic equilibrium; at $n \sim 10^{16}$ cm$^{-3}$, a Dark Star is born,
Most of the star is still hydrogen, with $<$ 1\% being DM, but this small amount is enough to provide a heat source for the star.

We have found the stellar structure and evolution of Dark Stars.  
DS are born with masses $\sim 1 \msun$.
They are giant puffy
(10 AU), cool (surface temperatures $\sim 10,000$K objects
(Freese et al. 2008a)). They reside in a large 
reservoir ($10^5 \msun$) of baryons, i.e. 15\% 
of the total halo mass. These baryons can start to accrete onto
the dark stars. Our work (Freese et al. 2008a; Spolyar et al. 2009) followed the
evolution of DS from their inception at $\sim 1 \msun$, as they accreted mass from the
surrounding halo.  

The key ingredient that allows dark stars to grow so much larger than ordinary fusion powered Population III stars is the fact that dark stars are so much cooler. Ordinary Pop III stars have much larger surface temperatures in excess of 50,000K. They produce ionizing photons that provide a variety of feedback mechanisms that cut off further accretion. McKee \& Tan (2008) have estimated that the resultant Pop III stellar masses are $\sim 140 \msun$. Dark stars are very different from fusion-powered stars, and their cooler surface temperatures allow continued accretion of baryons all the way up to enormous stellar masses, $M_*  > 10^5 \msun$.

Dark stars are stable as long as there is DM to fuel them.  There are two main methods for sustaining DM fuel:  (1)  gravitational attraction of dark matter particles on a variety of orbits (extended adiabatic contraction) and (2) capture by atomic nuclei due to elastic scattering. In our previous work we treated the DM halo as spherical and ran up the DS mass to the point where the DM initially inside the star was significantly reduced by annihilation;   this would limit the lifetime of the DS to $\sim 10^6$ yr.  
However, DM halos are triaxial, allowing for many kinds of orbits, including box orbits
and chaotic orbits, which can travel arbitrarily close to the center on any passage through the central cusp (Schwarzschild 1979, Goodman \& Schwarzschild 1981; Merritt \& Valluri 1996). Thus the cusp in a triaxial halo is constantly replenished by new orbits with low angular momentum. These orbits provide the steady state high density to allow DM particles bound to the cusps experience annihilation for a longer time.  This
extends the lifetime of DS, and can lead to larger and brighter DS's, including SMDS which could potentially lead to Supermassive Black Holes (Freese et at (2010)).

An additional source of DM is capture by atomic nuclei.  As DM from the halo passes through the DS, some WIMPs scatter off nuclei and are captured (Freese, Spolyar, Aguirre (2008); Iocco (2008)).  There are two uncertainties with this process, (1) the ambient DM density and (2) the scattering cross sections.  The scattering cross section is a free parameter, constrained only by bounds set by direct detection experiments.

\section{Detectability of Dark Stars in JWST and HST}

\begin{figure}
  \includegraphics[width=0.5\textwidth]{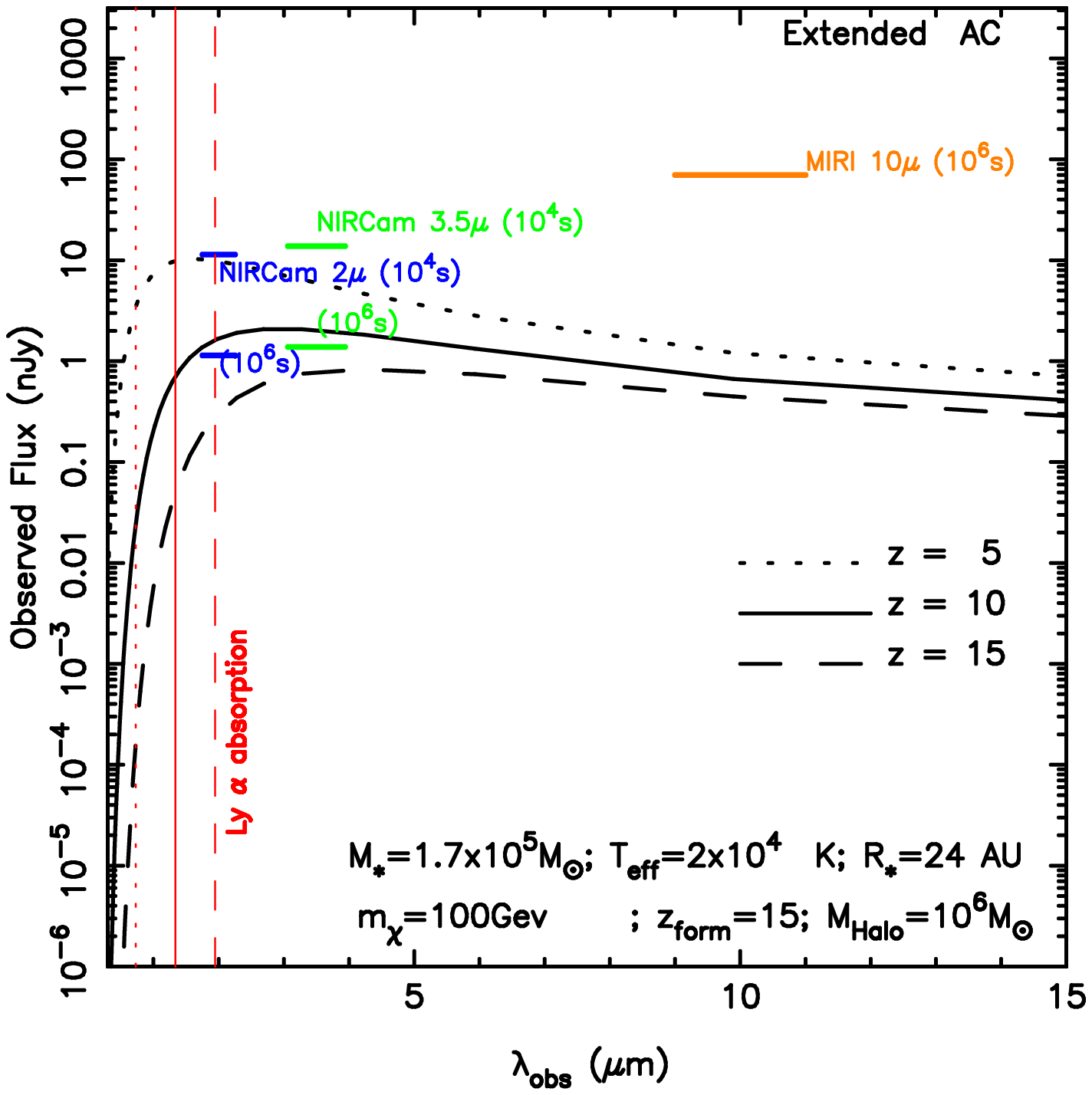}
  \includegraphics[width=0.5\textwidth]{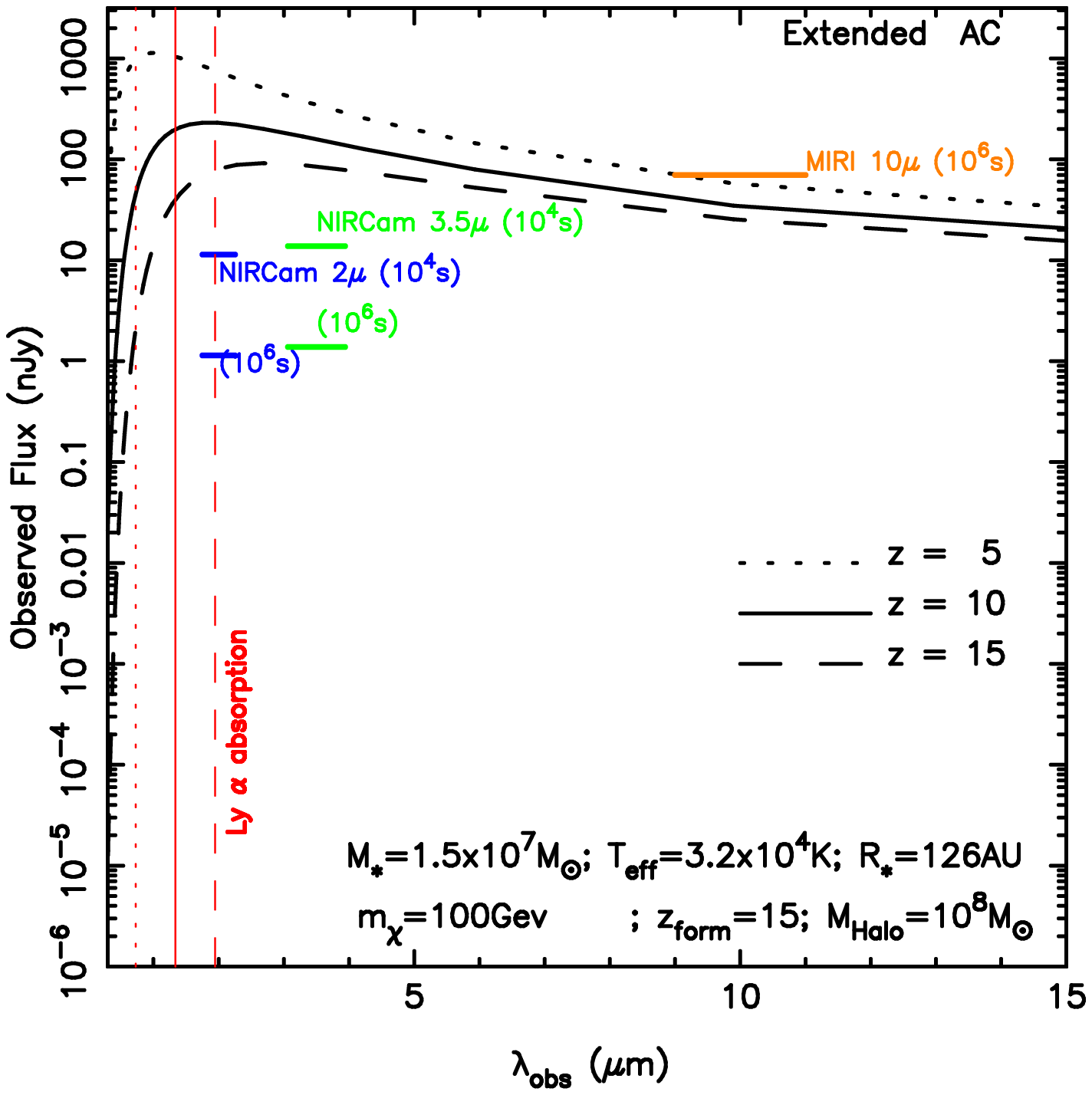}
  \caption{Detectability in JWST: Black body spectra of two dark stars formed via extended adiabatic contraction (``without capture'') for $\mwimp$=100 GeV. Left panel: $1.7\times 10^5~\msun$ SMDS in a $10^6~\msun$ halo. Right panel: $1.5\times 10^7~\msun$ SMDS in $10^8~\msun$ halo.  The black body flux is shown at $z=15$ (formation redshift) and at $z=10$ and 5 (see line legends) assuming that the dark star survives till the lower redshifts. Blue dashes show sensitivity limit and bandwidth of NIRCam $2\mu$ (R=4) while the green dashes show the sensitivity limit and band width of the NIRCam 3.5$\mu$ (R=4) band (Gardner et al, (2006,2009)). The upper and lower dashes show the sensitivity limits after exposure times of $10^4$s, $10^6$s respectively. The sensitivity of MIRI ($10\mu$, R=5) is shown for exposure time of $10^6$s (orange dash). All sensitivities are computed assuming a S/N=10.  The red vertical lines show the location of the 1216\AA line redshifted from the rest-frame wavelength of the star at each of the three redshifts. The observed flux to the left of the vertical lines will decrease relative to the black curves depending on the model assumed for IGM absorption up to the redshift of reionization.}
\label{jwst_flux_ac}
\end{figure}

SMDS can grow to be very bright: $10^5 \msun$ DS have luminosity $10^9 L_\odot$ while
$10^7 \msun$ DS can exceed $10^{11} L_\odot$.
Figure~\ref{jwst_flux_ac} shows the observed black body flux distribution of two SMDSs formed at $z=15$ for a WIMP mass $\mwimp = 100$GeV for the case of extended AC (without capture). The star in the left panel is formed in a $10^6\msun$ halo and the star in the right panel is formed in a $10^8~\msun$ halo and their stellar (baryonic) masses are $1.7\times 10^5~\msun$ and $1.5\times 10^7~\msun$ respectively.  Curves are shown assuming the SMDS formed at $z=15$ and survived to various redshifts, at which it is still producing blackbody radiation.  The $1.7\times 10^5~\msun$ star (left panel) will be detectable by JWST (NIRCam) in an exposure of a million seconds, but only if it survives intact till $z=10$.  The $1.5\times 10^7~\msun$ star (right panel) will be detectable even in a shorter $10^4s$ exposure even at $z=15$ in both the 2$\mu$ and 3.5$\mu$ bands. The star on the right may be marginally detectable in a million second exposure in the 10$\mu$ band of MIRI. The curves are not corrected for Ly-$\alpha$ absorption by the IGM but the red vertical lines show the location of the 1216\AA line redshifted from the rest-frame wavelength of the star at each of the three redshifts. Flux at wavelengths to the left of the redline at each redshift is expected to be absorbed to some extent by the IGM. 

\begin{figure}
  \includegraphics[width=.5\textwidth]{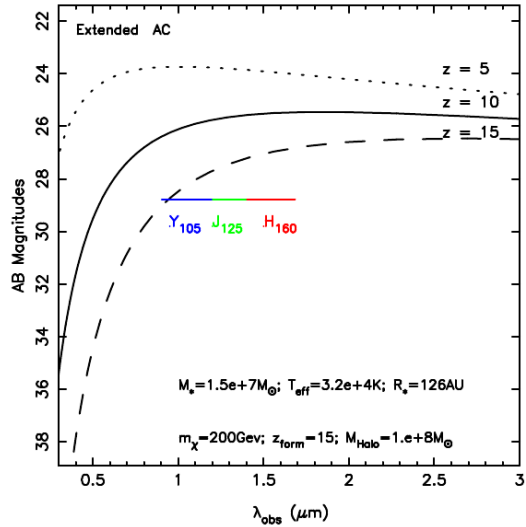}
  \caption{Detectability in HST: Similar to Fig.~\ref{jwst_flux_ac} for dark stars formed ``extended adiabatic contraction'' with $\mwimp=200$GeV as seen with HST.  Colored dashed show the sensitivity limits for the cameras on HST.}
\label{hst_flux_ac}
\end{figure}

Figure~\ref{hst_flux_ac} shows the detectability in HST of the
black body flux distribution of a $1.5 \times 10^7 \msun$ dark star formed in a DM halo $M_* = 10^8 \msun$ at $z=15$, using a WIMP mass $\mwimp = 200$GeV for the case of extended AC (without capture).  The figure shows that these SMDSs  may already be detectable in the recently obtained WFC3/IR image of the Hubble Ultra Deep Field (HST GO 11563:PI Illingworth) as $J_{125}$-band dropouts (Bouwens et al. 2010). 
Such a DS would be  distinguishable from  young star forming galaxies by future Spitzer IRAC observations taken as part of the GOODS program (PI:  M. Dickinson). Recently Bouwens et al. (2010b) reported the discovery of three possible $z\sim 10$ objects  in the WFC3/IR HUDF09. In this same HUDF image Bouwens et al. (2010a) discovered numerous sources at $z\sim 7$ and $z\sim 8$. The objects at $z\sim 7$ were also examined by Labb\'{e} et al (2010) and when co-added they were found to have significant increase (relative to the rest frame UV) in flux at 3.6$\mu$m and 4.5$\mu$m in deep Spitzer IRAC observations of this field obtained as part of the GOODs survey (Ouchi et al 2009).  The detections span the range of wavelengths from rest-frame UV to rest frame optical and are well fitted by standard Bruzual-Charlot type stellar population synthesis codes with fairly low metallicity (0.2$Z_\odot$) implying that the $z=7$ systems are star forming galaxies with populations as old as $\sim 350$~Myr. So far the $z=8-10$ objects are not detected at 3.6$\mu$m and 4.5$\mu$m by Spitzer IRAC  but continuing observations of this field as part of the GOODS program will undoubtedly be monitored. 

Since SMDS are significantly cooler than typical young hot stars in
the normal stellar populations at $z=10$, SMDS would not emit a
significant amount of flux in the UV at wavelengths shorter than the
Lyman alpha line ( see Fig. 1 {\it Right}). Consequently there would be little UV flux available to be reprocessed and emitted at longer wavelengths. Intrinsically, they are too faint to be detectable at longer wavelengths in a Spitzer IRAC image. This implies that SMDS
could be distinguished from normal stellar populations in young star forming galaxies in future WFC3/IR and Spitzer IRAC observation in two ways (1) they would be point
sources which are not associated with extended stellar distributions
(2) their flux at longer wavelengths would decrease steadily instead
of increasing as in the case of normal  stellar populations with ages $\sim100$Myr. 

Another paper on detectability of DS in JWST was done by Zackrisson et al (2010a).
They focused on lighter DS $\sim 10^3 \msun$, and pointed out that lensing may enhance the signal.

\section{Concluding Remarks}

This talk has focused on supermassive dark stars (SMDS) wtih  $M_* > 10^5\msun$. Such large masses are possible because the dark star is cool enough (as long as it is powered by DM) so that radiative feedback effects from the star do not shut off the accretion of baryons. These SMDS
are extremely bright (up to $10^{11} L_\odot$) and may thus be detectable in JWST.  In fact
they may have already been found in existing HST data or may appear in HST data taken
over the next year.

After a SMDS exhausts its DM fuel, it will have a short fusion phase, before eventually collapsing to a massive black hole.  SMDSs would make plausible precursors of the $10^9 \msun$ black holes observed at $z>6$ (Fan et al. 2003); of intermediate mass black holes; of BH at the centers of galaxies; and of the BH inferred by extragalactic radio excess seen by the ARCADE experiment (Seiffert et al. 2009).  In addition, the BH remnants from DS could play a role in high-redshift gamma ray bursts thought to take place due to accretion onto early black holes.

Subsequent to this presentation at the conference in Austin, a recent paper has examined in more detail
the detectability of DS in HST (Zackrisson et al (2010b)) and presented bounds on the numbers of
the most massive SMDS.  Their work has the advantage of using a stellar atmosphere code
rather than the far too simple blackbody approximation of our work.  More work of this type is in order.


\bibliographystyle{aipproc}   

\begin{thebibliography}{6}
\expandafter\ifx\csname natexlab\endcsname\relax\def\natexlab#1{#1}\fi
\providecommand{\enquote}[1]{``#1''}
\expandafter\ifx\csname url\endcsname\relax
  \def\url#1{\texttt{#1}}\fi
\expandafter\ifx\csname urlprefix\endcsname\relax\def\urlprefix{URL }\fi
\providecommand{\eprint}[2][]{\url{#2}}

\bibitem{FERMI}
  Abdo, A. A.,  et al. 2009  [The Fermi LAT Collaboration],
  Phys.\ Rev.\ Lett.\,    102, 181101 

\bibitem{PAMELA}
  Adriani, O., et al. 2009   [PAMELA Collaboration],
  Nature, 458, 607 

\bibitem{cdms}
  Ahmed, Z., et al. 2009  [The CDMS-II Collaboration],
  arXiv:0912.3592 [astro-ph.CO]

\bibitem{bbks_86}
{Bardeen} J.~M.,  {Bond} J.~R.,  {Kaiser} N.,    {Szalay} A.~S.,  1986, ApJ,
  304, 15

\bibitem{barkanaloeb}
  Barkana, R., \& Loeb, A. 2001,
  Phys.\ Rep.,   349, 125 

\bibitem{bar_efs_87}
{Barnes} J.,  \& {Efstathiou} G.,  1987, ApJ, 319, 575

\bibitem{dama}
  Bernabei, R.,   et al. 2009 [DAMA collaboration], 
  arXiv:0912.0660 [astro-ph.GA].
  
\bibitem{blumenthal}
  Blumenthal, G. R.,  Faber,  S. M., Flores,  R., \&   Primack, J. R. 1986, 
  ApJ,  301, 27 

\bibitem{bowens1}
Bouwens, R.J., et al. 2010a, ApJ, 708, L69

\bibitem{bowens2}
Bouwens, R.J., et al. 2010b, ApJ, 709, L133

 \bibitem{brommlarson}
  Bromm, V., \& Larson, R. B. 2004, 
  Annu.\ Rev.\ Astron.\ Astrophys.,   42, 79 

\bibitem{collar}
  Collar, J. I., \& McKinsey, D. N., 2010 arXiv:1005.0838v3 [astro-ph.CO]

\bibitem{fan_etal_03} Fan, X., et al.\ 2003, AJ
125, 1649 

\bibitem{Freese:2008wh}
  Freese,  K., Bodenheimer, P.,  Spolyar, D., \&  Gondolo, P. 2008a, 
  ApJ,   685, L101   

\bibitem{DS3}
  Freese, K.,  Gondolo, P., Sellwood, J. A., \& Spolyar, D. 2009, 
  ApJ,   693, 1563 

\bibitem{DMcap}
  Freese,  K., Spolyar, D., \&  Aguirre, A. 2008b, 
  J. Cosmol. Astropart. Phys., 
  JCAP,  0811, 014 

\bibitem{Freese:2010}
  Freese,  K., Ilie,  C., Spolyar,  D., Valluri,  M. \& Bodenheimer,  P., 2010, arxiv:1002.2233v1

\bibitem{gardner_etal_06} Gardner, J.~P., et al.\ 
2006, Space Science Reviews, 123, 485 


\bibitem{gardner_etal_09a} Gardner, J.~P., et al.\ 
2009, Astrophysics in the Next Decade, Astrophysics and Space Science 
Proceedings, Volume .~ISBN 978-1-4020-9456-9.~Springer Netherlands, 2009, 
p.~1, 1 
 
 \bibitem{good_schwarz_81} Goodman, J., \& Schwarzschild, M.\ 1981, ApJ, 245, 1087

\bibitem{hooper}
Hooper, D. , Finkbeiner, D. , \& Gregory Dobler, G., 2007, Phys. Rev. D 76, 083012
 
\bibitem{iocco}
  Iocco, F. 2008,  ApJ, 667, L1 

\bibitem{iocco2}
Iocco, F.,  Bressan, A.,  Ripamonti, E.,  Schneider, R., Ferrara, A., 
\& Marigo, P. 2008,  MNRAS, 390, 1655

\bibitem{labbe}
Labb\'{e} , I., et al. 2010a, ApJ, 708, L26

\bibitem{mckeetan}
  McKee, C. F., \&  Tan, J. C. 2008, 
  ApJ, 681, 771
 
\bibitem{merritt_valluri_96}
Merritt, D.,  \&  Valluri, M. 1996, ApJ, 471, 82

\bibitem{ouchi}
Ouchi, M. et al 2009, ApJ 706 1136

\bibitem{ripamontiabel}
  Ripamonti, E., \& Abel, T. 2005, 
  arXiv:astro-ph/0507130.

\bibitem{ripamontione}
Ripamonti, E., Iocco,, F., Bressan, A., Schneider, R., Ferrara, A.,  and Marigo, P., 2009,
  PoS {\bf IDM2008}, 075

\bibitem{ripamontietal}
Ripamonti, E., et al 2010, arXiv:1003.0676v1

\bibitem{schwarz_79} Schwarzschild, M.\ 1979, 
ApJ, 232, 236 

\bibitem{seiffert}
 Seiffert, M.,   et al. 2009, 
  arXiv:0901.0559 [astro-ph.CO].

\bibitem{DSnl}
  Spolyar, D.,  Bodenheimer, P., Freese, K., \&  Gondolo, P. 2009, 
  ApJ,   705, 1031 

\bibitem{DS2}
 Spolyar, D.,  Freese, K., \&  Gondolo, P. 2008, 
  Phys.\ Rev.\ Lett.\,  100, 051101 

\bibitem{umeda}
Umeda,  H., Yoshida, N.,  Nomoto, K.,  Tsuruta, S.,  Sasaki, M., 
\& Ohkubo, T. 2009, 
  J. Cosmol. Astropart. Phys., 
  JCAP, 0908, 024 

\bibitem{yoshida03}
Yoshida, N.,  Abel, T.,  Hernquist ,L.,  \& Sugiyama, N. 2003, 
ApJ,  592, 645 

\bibitem{zackrisson1}
Zackrisson, E., Scott, P., Rydberg, C. E., Iocco, F., Edvardsson, B., \"{O}stlin, G., Sivertsson, S., Zitrin, A., Broadhurst, T., Gondolo, P. 2010a, ApJ 717, 257,  arXiv:1002.3368v2 [astro-ph.CO]

\bibitem{zackrisson2}
Zackrison, E., Scott, P., Rydberg, C. E., Iocco, F., Silvertsson, S., \"{O}stlin, G.,  Mellema, G., Iliev, I., Shapiro, P., 2010b, arxiv: 1006.0481v1 [astro-ph.CO]

\end{thebibliography}

\hyphenation{Post-Script Sprin-ger}

\end{document}